\documentclass{ws-procs975x65}
\newcommand{\figcaption}[1]{\def\@captype{figure}\caption{#1}}
\newcommand{\tblcaption}[1]{\def\@captype{table}\caption{#1}}

\usepackage{amssymb}
\usepackage{amsmath}
\usepackage{epsfig}
\usepackage{graphicx}
\usepackage[dvips]{color}

\def\simge{\mathrel{%
       \rlap{\raise 0.511ex \hbox{$>$}}{\lower 0.511ex \hbox{$\sim$}}}}
\def\simle{\mathrel{
       \rlap{\raise 0.511ex \hbox{$<$}}{\lower 0.511ex \hbox{$\sim$}}}}
\begin{document}

\title{
Phase structure of many flavor lattice QCD at finite temperature
}
\author{Norikazu Yamada\footnote{Speaker. E-mail: norikazu.yamada@kek.jp}}
\address{
KEK Theory Center, Institute of Particle and Nuclear Studies, High
Energy Accelerator Research Organization (KEK), Tsukuba 305-0801,
Japan\\
School of High Energy Accelerator Science, The Graduate University
for Advanced Studies (Sokendai), Tsukuba 305-0801, Japan
}
\author{Shinji Ejiri}
\address{
Graduate School of Science and Technology, Niigata University, Niigata
950-2181, Japan
}
\begin{abstract}
 In realistic technicolor models containing many fermions, the
 electroweak baryogenesis offers a natural scenario for generating
 baryon number asymmetry.
 One of the key ingredients is the occurrence of the first order phase 
 transition at finite temperature.
 As a first step toward the exploration of this possibility on the
 lattice, we develop an agile method to identify the critical mass for a
 given $N_f$, separating the first order and the crossover transition.
 We explain the outline of our method and demonstrate it by determining the
 critical mass of $N_f$-flavors in the presence of light two-flavors.
 It is found that the critical mass becomes larger with $N_f$.
\end{abstract}

\keywords{Lattice gauge theory, Technicolor models, quark-gluon plasma.}
\bodymatter

\section{Introduction}

Technicolor (TC)~\cite{Weinberg:1975gm} has been known as one of the
most natural candidates for physics beyond the standard model.
The classical TC models have been already excluded for many reasons,
while those with many fermion flavors in the fundamental representation
are expected to escape from various experimental constraints due to
special properties called ``walking'' dynamics~\cite{Holdom:1981rm}.
Whether many-flavor TC models really show expected walking behaviors is
now actively and rigorously investigated on the
lattice~\cite{DelDebbio:2010zz}.

In this article, we focus on another aspect of TC models, the nature of
the thermal phase transition.
We consider as a TC model SU(3) gauge theory with 2+$N_f$ flavors of
techniquarks where the mass of two-flavors are fixed to a small value
and the other $N_f$ flavors have arbitrary masses.
$N_f$ is also taken to be arbitrary.
Based on an analysis of linear sigma
model~\cite{Pisarski:1983ms}\footnote{As for an interesting
argument about the phase transition in two-flavor QCD, see
Ref.~\refcite{Aoki:2012yj}.}, we infer that the nature of the chiral phase
transition at finite temperature changes from crossover to first order
at a critical mass as the mass of $N_f$ flavors are decreased from
infinity while keeping the two-flavors' mass constant.
It should be noted that, if the first order transition is strong enough
(or equivalently the masses of $N_f$-flavors are light enough), the
electroweak baryogenesis (EWBG) would become viable.
EWBG is, in general, restrictive and attractive in that no tunable
parameter exists and its success is solely determined by the dynamics
intrinsic to gauge theory.
Then, it is interesting to ask what is the upper bound on the masses of
$N_f$-flavors which allows the first order phase transition and
successful EWBG.
EWBG in the standard model (SM) was studied on the lattice and turned
out to fail since $m_H\sim$ 125 GeV is too heavy to induce the first
order transition~\cite{Fodor:1999at}.
As for TC models, the studies in this direction were carried out in the
context of the effective theory and obtained promising
results~\cite{Appelquist:1995en}.
As a first step toward the rigorous test of this possibility we study
the thermal nature of a many-flavor TC model by lattice numerical
simulations.
To be precise, we aim at putting the upper bound on the mass of
$N_f$-flavors of fermions by requiring the occurrence of first order
phase transition.
This upper bound can then be translated into that on the technipion
mass, which can be directly compared to the results of LHC.

Here let us describe why we consider ``2+$N_f$''-flavors.
In many-flavor TC models without the $SU_L(2)\times U_Y(1)$
interactions, two flavors of them have to be exactly massless and the
resulting three massless Nambu-Goldstone bosons (NGBs) are absorbed into
the longitudinal mode of the weak gauge bosons when one turns on the
$SU_L(2)\times U_Y(1)$ interactions.
On the other hand, the mass of other $N_f$ flavors must be larger than
an appropriate lower bound otherwise S$\chi$SB produces too many (light
pseudo) NGBs, none of which is observed yet.
Furthermore, in the presence of too many massless and almost massless
NGBs, $S$-parameter~\cite{Peskin:1990zt,Shintani:2008qe} becomes large
or even diverges.
Thus, $N_f$-flavors have to have explicit breakings of an appropriate
size.
In this work, we simply let them have a mass.

Our model based on SU(3) gauge theory is essentially the same as many
flavor QCD except for their dynamical scales; $\sim$1 TeV for TC and
$\sim$1 GeV for QCD and thus we can simply apply numerical techniques
developed in lattice QCD to the study of TC.
As discussed below, the critical mass increases with $N_f$.
Hence, from the viewpoint of lattice numerical simulation, the boundary
of the first order region can be reached more easily for large $N_f$.

Another purpose of this study is to understand the real QCD with 2+1
flavors.
At the physical masses and zero density, the chiral transition is known
to be crossover, and is expected to become first order at a critical
density.
Toward the determination of the critical density, it is important to
find the critical surface in the parameter space spanned by masses and
chemical potential~\cite{dFP,crtpt}.
However, recent lattice QCD studies suggest that the critical surface at 
zero density is located in the very light quark mass region and it makes
the determination extremely difficult~\cite{RBCBi09}.
Fortunately, some of properties are independent of $N_f$.
The study of 2+$N_f$-flavor QCD is expected to provide important
information for 2+1-flavor QCD.

We first describe the method to identify the nature of the phase
transition and then present the critical mass separating
the first order and crossover regions in 2+$N_f$-flavor QCD.
The work reported here has been already published in
Ref.~\refcite{Ejiri:2012rr}.

\section{Method}

We examine the effective potential defined by the probability
distribution function of the gauge action to identify the nature
of the phase transition.
The first order transition is concluded by the existence of the two
peaks in the distribution function~\cite{Ejiri:2007ga,whot11}.
We define the distribution function for 2+$N_f$-flavor QCD with the
quark masses $m_f$ ($f=1, \cdots, 2+N_f$) by
\begin{eqnarray}
\vspace*{-15ex}
w(P; \beta, m_f) 
&=&  \int {\cal D} U {\cal D} \psi {\cal D} \bar{\psi} \
    \delta(P- \hat{P}) \ e^{- S_q - S_g} \nonumber
\\
\vspace{-15ex}
&=& \int {\cal D} U \ \delta(P- \hat{P}) \ 
    e^{6\beta N_{\rm site} \hat{P}}\
    \prod_{f=1}^{N_f+2} (\det M(m_f)),
\vspace{-15ex}
\label{eq:pdist}
\end{eqnarray}
where 
$S_g$ and $S_q$ are the gauge and quark actions, respectively, and $M$
is the quark matrix.
$N_{\rm site} \equiv N_{\rm s}^3 \times N_t$ is the number of sites.
$\beta=6/g_0^2$ is the inverse lattice bare coupling, and
$\hat P=-S_g/(6N_{\rm site} \beta)$.
The effective potential is then defined by
\begin{eqnarray}
 V_{\rm eff}(P;\beta,m_f) = -\ln w(P;\beta,m_f).
 \label{eq:effective-potential}
\end{eqnarray}

We consider QCD with two degenerate light quarks of the mass
$m_{\rm l}$ and $N_f$ quarks of $m_h$.
For later convenience, the potential is separated into two parts; one is
the contribution from two-flavor QCD $V_0(P; \beta)$ and the other is the
rest,
\begin{eqnarray}
    V_{\rm eff}(P; \beta, m_h)
=   V_0(P; \beta_0) - \ln R(P; \beta, m_h; \beta_0),
\label{eq:vefftrans}
\end{eqnarray}
with
\begin{eqnarray}
\ln R(P; \beta, m_h; \beta_0)
&=& 6(\beta - \beta_0)N_{\rm site}P 
 + \ln
     \left\langle
     \displaystyle
     \prod_{h=1}^{N_f}
     \frac{\det M(m_h)}{\det M(\infty)}
     \right\rangle_{P: {\rm fixed}}, \ 
\label{eq:lnr}
\end{eqnarray}
where 
$ \langle \cdots \rangle_{P: {\rm fixed}} \equiv 
\langle \delta(P- \hat{P}) \cdots \rangle_{\beta_0} /
\langle \delta(P- \hat{P}) \rangle_{\beta_0} $
and $\langle \cdots \rangle_{\beta_0}$ denotes the ensemble average over
two-flavor configurations generated at $\beta_0$ and $m_l$.
Since the $m_l$ dependence is not discussed in the following, it is
omitted from the arguments.
$\beta_0$ is the simulation point, which may differ from $\beta$.
By performing simulations at various $\beta_0$, one can obtain the
potential in a wide range of $P$.

Restricting the calculation to the heavy quark region, the
determinant for $N_f$ flavors in eq.~(\ref{eq:lnr}) is approximated at
the leading order as
\begin{eqnarray}
\ln \left[ \frac{\det M (\kappa_h)}{\det M (0)} \right]
=  288 N_{\rm site} \kappa_h^4 \hat{P}
 + 12 N_s^3 (2 \kappa_h)^{N_t} \hat{\Omega}
+ \! \! \cdots
\label{eq:detmw}
\end{eqnarray}
for the standard Wilson quark action and
\begin{eqnarray}
\ln \left[ \frac{\det M (m_{\rm h})}{\det M (\infty)} \right]
=  \frac{36 N_{\rm site}}{(2m_{\rm h})^4} \hat{P}
+  \frac{6 N_s^3}{(2m_{\rm h})^{N_t}} \hat{\Omega}
+ \cdots
\label{eq:detms}
\end{eqnarray}
for the four-flavor standard staggered quark with $m_{\rm h}$.
$\kappa_h$ in eq.~(\ref{eq:detmw}) is the hopping parameter being 
proportional to $1/m_h$, and $\hat{\Omega}$ is the real part of the
Polyakov loop.
For improved gauge actions such as
$S_g = 6N_{\rm site} \beta [c_0 {\rm (plaquette)}
+ c_1 {\rm (rectangle)}]$,
additional $c_1\times O(\kappa^4)$ terms must be contained
in eqs.~(\ref{eq:detmw}) and (\ref{eq:detms}),
where $c_1$ is the improvement coefficient and $c_0=1-8c_1$.
However, since the improvement term does not affect the physics,
we will cancel these terms by a shift of $c_1$.

Beyond the critical $\beta$ corresponding to the endpoint of a first order
transition, $V_{\rm eff}$ takes a double-well shape as a function
of $P$, and equivalently the curvature of the potential
$d^2 V_{\rm eff}/d^2P$ will take a negative value.
Observing this behavior usually requires a fine-tuning of $\beta$.
However, $d^2V_{\rm eff}/dP^2$ is independent of $\beta$ and
$d^2V_{\rm eff}/dP^2$ over the wide range of $P$ can be easily obtained
by combining data obtained at different $\beta$.
Thus the fine-tuning of $\beta$ is not necessary in this
case~\cite{Ejiri:2007ga}.
We focus on the curvature of the effective potential to
identify the nature of the phase transition.

Denoting $h=2 N_f (2\kappa_h)^{N_t}$ for $N_f$ degenerate Wilson quarks, 
or $h=N_f/(4 \times (2m_{\rm h})^{N_t})$ for the staggered quarks, 
we obtain
\begin{eqnarray}
    \ln R(P;\beta,\kappa_h;\beta_0)
&=& \ln\bar{R}(P; h)+{\rm (plaquette \ term)} + O(\kappa_h^{N_t+2})\\
    \bar{R}(P; h)
&=& \left\langle \exp [6h N_s^3 \hat{\Omega}]
\right\rangle_{P: {\rm fixed}, \beta_0}.
\label{eq:rew2f}
\end{eqnarray}
Notice that $\bar{R}(P; h)$ is independent of $\beta_0$.
The plaquette term does not contribute to $d^2V_{\rm eff}/dP^2$ and can
be absorbed by shifting
$\beta \to \beta^{*} \equiv \beta + 48 N_f \kappa_h^4$ for Wilson
quarks.
$N_f$-flavors do not have to be degenerate.
The non-degenerate case is realized by redefining
$h=2 \sum_{f=1}^{N_f} (2 \kappa_f)^{N_t}$ or
$h=(1/4)\sum_{f=1}^{N_f} (2m_f)^{-N_t}$.
In the following, we discuss the mass dependence of $\bar{R}$ through
the parameter $h$.

\begin{figure}[tb]
\begin{center}
\hspace*{-5ex}
\begin{tabular}{cc}
 \includegraphics[width=0.5 \textwidth]{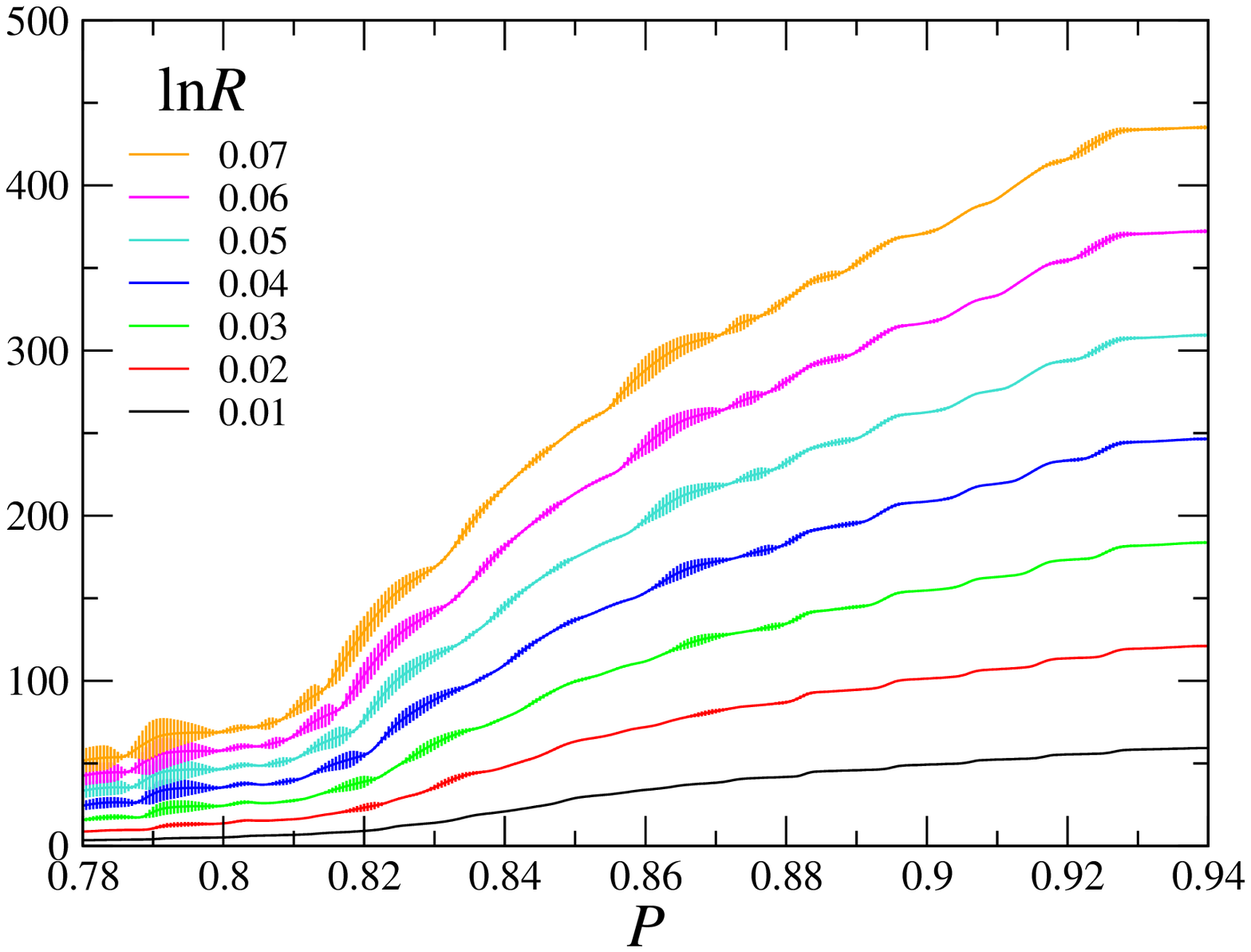} &
 \includegraphics[width=0.5 \textwidth]{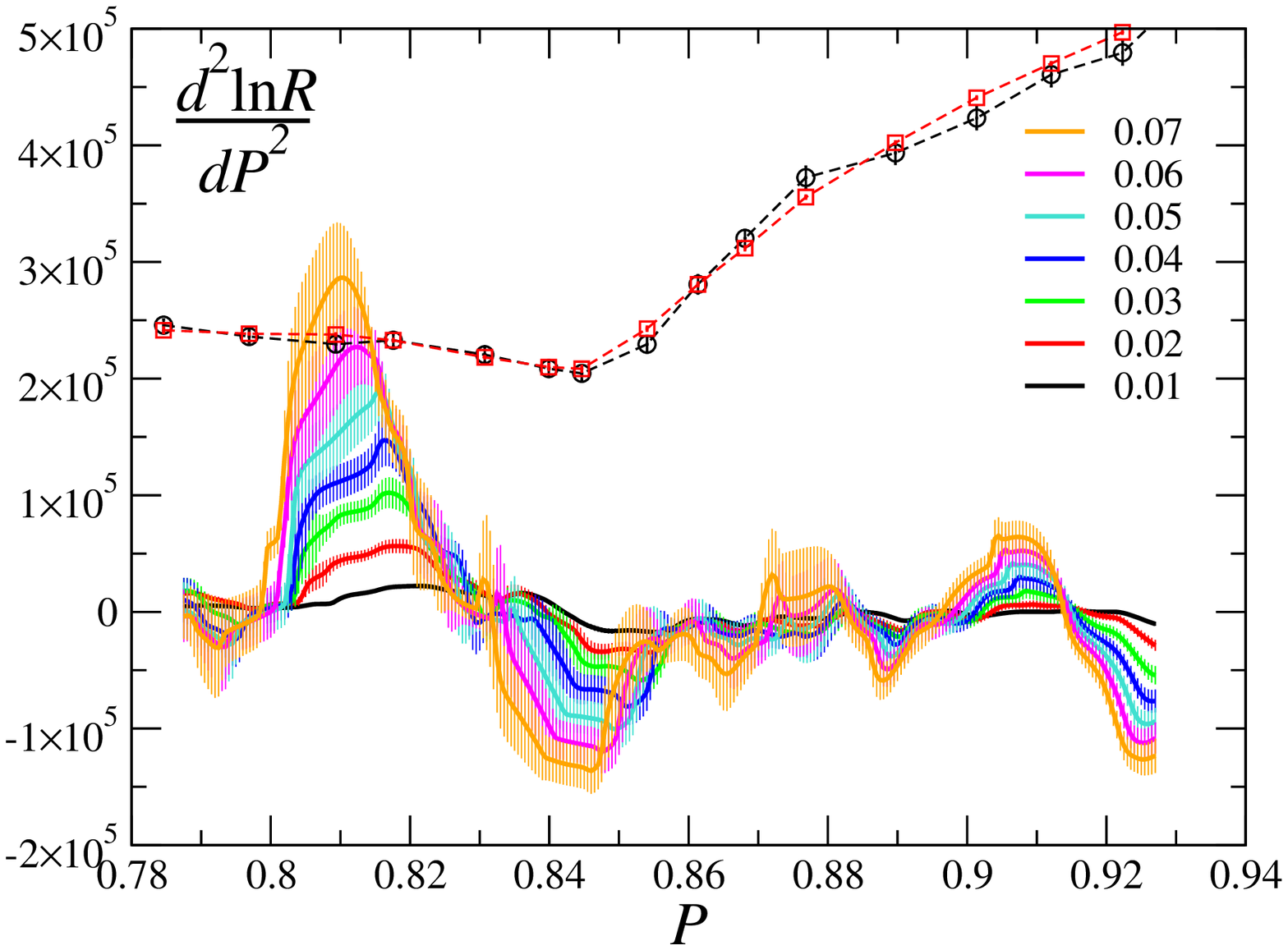}
\end{tabular}
\caption{Left: $\ln \bar{R} (P; h,0)$ as functions of the gauge action.
Right: The curvature of $\ln \bar{R} (P; h,0)$ for $h=0.01$ -- $0.07$.
The circle and square symbols are $d^2V_0/dP^2(P)$.}
\label{fig:lnr}
\end{center}
\end{figure}

\section{Numerical results}

We explicitly demonstrate the above method.
We use the two-flavor QCD configurations generated with p4-improved
staggered quark and Symanzik-improved gauge actions in Ref.~\refcite{BS05}.
The lattice size $N_{\rm site}$ is $16^3 \times 4$, and the data are
obtained at sixteen values of $\beta$ from $\beta=3.52$ to $4.00$
keeping the bare quark mass to $ma=0.1$.
The number of trajectories is 10,000 -- 40,000, depending on $\beta$.
The corresponding temperature normalized by the pseudo-critical
temperature is in the range of $T/T_c= 0.76$ to $1.98$, and the
pseudo-critical point is about $\beta=3.65$, where the $\pi$-$\rho$
ratio is $m_{\rm PS}/m_{\rm V} \approx 0.7$.
Further details on the simulation parameters are given in
Ref.~\refcite{BS05}.
The same data set is used to study the phase structure of two-flavor QCD
at finite density in Ref.~\refcite{Ejiri:2007ga}. 

\begin{figure}[tb]
\begin{center}
\begin{tabular}{c}
\includegraphics[width=71mm]{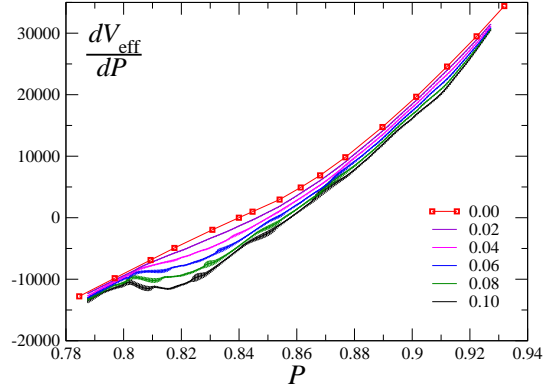}
\end{tabular}
\vspace{-3mm}
\caption{The 
slope of $V_{\rm eff} (P; \beta, h,0)$ 
normalized at $(\beta, h)=(3.65, 0)$ for $h=0.0$ -- $0.1$.
The squares are $dV_0/d P$.}
\label{fig:vslp}
\end{center}
\end{figure}

We first calculate the potential in two-flavor QCD, $V_0(P; \beta)$, the
first term in eq.~(\ref{eq:vefftrans}).
Because the finite temperature transition is crossover for two-flavor
QCD at a finite quark mass, the distribution function is always Gaussian
type.
We thus evaluate the curvature of $V_0$ using an identity for the
Gaussian distribution, $d^2 V_0/dP^2 = 6N_{\rm site}/\chi_P$, where
$\chi_P$ is the gauge action susceptibility,
\begin{eqnarray}
    \chi_P
\equiv
    6 N_{\rm site} \langle (P- \langle P \rangle )^2 \rangle. 
\end{eqnarray}
The slope of $V_0$ in the heavy quark limit can be also measured using
an equation derived from eqs.~(\ref{eq:vefftrans}) and (\ref{eq:lnr}).
When one performs a simulation at $\beta_0$, the slope is zero at the
minimum of $V_0(P; \beta_0)$, and the minimum is realized at
$P \approx \langle \hat{P} \rangle_{\beta_0}$. 
Hence, we obtain~\cite{ejiri09}
\begin{eqnarray}
  \frac{d\,V_0( \langle \hat{P} \rangle_{\beta_0}, \beta)}{d\,P} 
= - 6(\beta - \beta_0)N_{\rm site}.
\end{eqnarray}
The result of $d^2V_0/dP^2$ is plotted in the right panel of
Fig.~\ref{fig:lnr}.
The circles with dashed lines are calculated by $\chi_P$. 
The squares are computed by the numerical differential of $dV_0/dP$
obtained at the minimum of $V_0$.
$dV_0/dP$ are the squares in Fig.~\ref{fig:vslp}.
It is seen that two different methods provide the consistent results.

In the calculation of $\bar{R}(P; h)$, we use the delta function
approximated by
$\delta(x) \approx 1/(\Delta \sqrt{\pi}) \exp[-(x/\Delta)^2]$,
where $\Delta=0.0025$ is adopted consulting the resolution and the
statistical error.
Because $\bar{R}(P; h)$ is independent of $\beta$, the data obtained at
various $\beta$ are gathered as is done in Ref.~\refcite{Ejiri:2007ga}.
The results for $\ln \bar R(P;h)$ are shown by solid curves in the left
panel of Fig.~\ref{fig:lnr} for $h=0.01$ -- $0.07$.
A rapid increase is observed around $P \sim 0.82$.
It is also important to note that the gradient becomes larger with $h$.

The second derivative $d^2 \ln \bar R/dP^2$ is calculated by fitting
$\ln \bar R$ to a quadratic function of $P$ with a range of
$P \pm 0.015$ and repeating with various $P$.
The results are plotted in Fig.~\ref{fig:lnr} (right), where
$d^2V_0/dP^2$ is also shown as the circles or the squares with dashed
lines.
This figure shows that $d^2 (\ln \bar R)/dP^2$ becomes larger with $h$,
and the maximum around $P=0.81$ exceeds $d^2 V_0/dP^2$ for $h > 0.06$.
This indicates that the curvature of the effective potential,
$d^2 V_{\rm eff}/dP^2 =d^2 V_0 /dP^2 -d^2 (\ln \bar R)/dP^2$, vanishes
at $h\sim 0.06$ and for large $h$ there exists a region of $P$ where the
curvature is negative.
We estimated the critical value $h_c$ at which the minimum of
$d^2 V_{\rm eff}/dP^2$ vanishes and obtained $h_c=0.0614(69)$.

To see the appearance of the first order transition in a different way,
we show $dV_{\rm eff}/dP$ at finite $h$ for $\beta^*=3.65$ in
Fig.~\ref{fig:vslp}.
The shape of the $dV_{\rm eff}/dP$ is independent of $\beta$ because
$d^2 V_{\rm eff}/dP^2$ is $\beta$-independent.
$dV_{\rm eff}/dP$ monotonically increases when $h$ becomes small,
indicating that the transition is crossover.
However, the shape of $dV_{\rm eff}/dP$ turns into an S-shape at
$h \sim 0.06$, corresponding to the double-well potential.

We defined the parameter $h=2 N_f \times (2\kappa_h)^{N_t}$ for the
Wilson quark.
Then, the critical $\kappa_{hc}$ corresponding $h_c$ decreases as
$\kappa_{hc}=[h_c/(2N_f)]^{1/N_t}/2$ with $N_f$, and the truncation
error from the higher order terms in $\kappa_h$ becomes smaller as $N_f$
increases.
The application range of the hopping parameter expansion was examined in
quenched QCD simulations with $N_t=4$, by explicitly measuring the size
of the next-to-leading order (NLO) terms of the expansion~\cite{whot12}.
Then the NLO contribution turned out to be comparable to that in the
leading order at $\kappa_h \sim 0.18$.
Hence, this method may be applicable up to around $\kappa_h\sim 0.1$.
For instance, in the case of $N_f=10$ with $N_t=4$,  $\kappa_{hc}$ is
0.118.

\section{Conclusion and outlook}

We proposed an agile method to study the thermal nature of many-flavor
QCD with EWBG in TC in mind, and applied it to the 2+$N_f$-flavor QCD.
Fixing the mass of two light quarks, we determined the critical mass of
the remaining $N_f$-flavors, which separates the first order and
crossover regions.
The critical mass is found to become larger with $N_f$.
Further studies using this method are given in Ref.~\refcite{Ejiri:2012rr},
including the investigations at finite density.
We find that the critical mass increases with $\mu$ in the
2+$N_f$-flavor QCD.

The next step for the estimation of the baryon number asymmetry in TC
scenario is to quantify the strength of the first order phase
transition.
Another interesting application of our method is to study universal
scaling behavior near the tricritical point.
If the chiral phase transition in the two flavor massless limit is of
second order, the boundary of the first order transition region
$m_{\rm l}^c (m_{\rm h})$ is expected to behave as
$m_{\rm l}^c \sim |m_h^{\rm tri.} -m_h|^{5/2}$ in the vicinity of the
tricritical point,
$(m_{\rm l}, m_{\rm h}, \mu)=(0, m_{\rm h}^{\rm tri.}, 0)$,
from the mean field analysis.
This power behavior is universal for any $N_f$. 
The density dependence is important as well, which is expected to be
$m_{\rm l}^c \sim |\mu|^5$ \cite{ejirilat08}.
Starting from large $N_f$, the systematic study of properties of real
QCD phase transition is possible.

\section*{Acknowledgments}
We would like to thank the organizers of this fruitful workshop.
A part of this work was completed at GGI work shop.
This work is in part supported by Grants-in-Aid of the Japanese Ministry
of Education, Culture, Sports, Science and Technology
(No.\
22740183, 
23540295 
)
and by the Grant-in-Aid for Scientific Research on Innovative Areas
(No.\
20105002, 
20105005, 
23105706  
).

\end{document}